\pdfoutput=1
\documentclass[12pt,a4paper]{article}

\usepackage{ifthen} 
\newboolean{pdflatex}
\setboolean{pdflatex}{true} 

\newboolean{articletitles}
\setboolean{articletitles}{true} 

\newboolean{uprightparticles}
\setboolean{uprightparticles}{false} 

\newboolean{inbibliography}
\setboolean{inbibliography}{false} 

\def\paperauthors{The Heavy Flavor Averaging Group (HFLAV)} 
\def\paperasciititle{HFLAV input to the update of the European Strategy for Particle Physics} 
\def\papertitle{HFLAV input to the update of the \\ European Strategy for Particle Physics} 
\def\paperkeywords{{High Energy Physics}, {HFLAV}, {flavor physics}} 
\def\papercopyright{2018 HFLAV} 
\def\paperlicence{CC-BY-4.0 licence}
\def\paperlicenceurl{https://creativecommons.org/licenses/by/4.0/}

\usepackage[top=1in, bottom=1.25in, left=1in, right=1in]{geometry}

\usepackage[utf8]{inputenc}
\usepackage[T1]{fontenc}
\usepackage{microtype}
\usepackage{lineno}  
\usepackage{xspace} 
\usepackage{caption} 

\usepackage{graphicx}  
\usepackage{color}
\usepackage{colortbl}
\graphicspath{{./figs/}} 
\DeclareGraphicsExtensions{.pdf,.PDF,png,.PNG}

\usepackage{amsmath} 
\usepackage{amssymb}
\usepackage{amsfonts}
\usepackage{upgreek} 

\newcommand*\patchAmsMathEnvironmentForLineno[1]{%
\expandafter\let\csname old#1\expandafter\endcsname\csname #1\endcsname
\expandafter\let\csname oldend#1\expandafter\endcsname\csname
end#1\endcsname
 \renewenvironment{#1}%
   {\linenomath\csname old#1\endcsname}%
   {\csname oldend#1\endcsname\endlinenomath}%
}
\newcommand*\patchBothAmsMathEnvironmentsForLineno[1]{%
  \patchAmsMathEnvironmentForLineno{#1}%
  \patchAmsMathEnvironmentForLineno{#1*}%
}
\AtBeginDocument{%
\patchBothAmsMathEnvironmentsForLineno{equation}%
\patchBothAmsMathEnvironmentsForLineno{align}%
\patchBothAmsMathEnvironmentsForLineno{flalign}%
\patchBothAmsMathEnvironmentsForLineno{alignat}%
\patchBothAmsMathEnvironmentsForLineno{gather}%
\patchBothAmsMathEnvironmentsForLineno{multline}%
\patchBothAmsMathEnvironmentsForLineno{eqnarray}%
}

\usepackage{hyperxmp}

\usepackage[pdftex,
            pdfauthor={\paperauthors},
            pdftitle={\paperasciititle},
            pdfkeywords={\paperkeywords},
            pdfcopyright={Copyright (C) \papercopyright},
            pdflicenseurl={\paperlicenceurl}]{hyperref}

\usepackage[colorinlistoftodos,textsize=scriptsize]{todonotes}

\usepackage[all]{hypcap} 
\usepackage{authblk}

\usepackage{xspace} 
\usepackage{upgreek}

\newcommand{\offsetoverline}[2][0.1em]{\kern #1\overline{\kern -#1 #2}}%

\def\lhcb   {\mbox{LHCb}\xspace}
\def\atlas  {\mbox{ATLAS}\xspace}
\def\cms    {\mbox{CMS}\xspace}

\def\belle  {\mbox{Belle}\xspace}
\def\belletwo {\mbox{Belle~II}\xspace}
\def\besiii {\mbox{BESIII}\xspace}

\def\MagUp {\mbox{\em Mag\kern -0.05em Up}\xspace}

\ifthenelse{\boolean{uprightparticles}}%
{

 \def\PDelta      {\ensuremath{\Delta}\xspace}                 
 \def\PXi         {\ensuremath{\Xi}\xspace}                 
 \def\PLambda     {\ensuremath{\Lambda}\xspace}                 
 \def\PSigma      {\ensuremath{\Sigma}\xspace}                 
 \def\POmega      {\ensuremath{\Omega}\xspace}                 
 \def\PUpsilon    {\ensuremath{\Upsilon}\xspace}

 \def\PB      {\ensuremath{\mathrm{B}}\xspace}                 
                  
 \def\PD      {\ensuremath{\mathrm{D}}\xspace}

 \def\PK      {\ensuremath{\mathrm{K}}\xspace}

 \def\Pi      {\ensuremath{\mathrm{i}}\xspace}

}
{

 \mathchardef\PDelta="7101
 \mathchardef\PXi="7104
 \mathchardef\PLambda="7103
 \mathchardef\PSigma="7106
 \mathchardef\POmega="710A
 \mathchardef\PUpsilon="7107
                  
 \def\PB      {\ensuremath{B}\xspace}                 
                  
 \def\PD      {\ensuremath{D}\xspace}

 \def\PK      {\ensuremath{K}\xspace}

 \def\Pi      {\ensuremath{i}\xspace}

}

\makeatletter
\ifcase \@ptsize \relax
  \newcommand{\miniscule}{\@setfontsize\miniscule{4}{5}}
\or
  \newcommand{\miniscule}{\@setfontsize\miniscule{5}{6}}
\or
  \newcommand{\miniscule}{\@setfontsize\miniscule{5}{6}}
\fi
\makeatother

\DeclareRobustCommand{\optbar}[1]{\shortstack{{\miniscule (\rule[.5ex]{1.25em}{.18mm})}
  \\ [-.7ex] $#1$}}

  \def\Kbar    {{\kern 0.2em\overline{\kern -0.2em \PK}{}}\xspace}

\def\KorKbar {\kern 0.18em\optbar{\kern -0.18em K}{}\xspace}

  \def\Dbar    {{\kern 0.2em\overline{\kern -0.2em \PD}{}}\xspace}

\def\DorDbar {\kern 0.18em\optbar{\kern -0.18em D}{}\xspace}

\def\Bbar    {{\ensuremath{\kern 0.18em\overline{\kern -0.18em \PB}{}}}\xspace}

\def\BorBbar    {\kern 0.18em\optbar{\kern -0.18em B}{}\xspace}

\def\Y#1S{\ensuremath{\PUpsilon{(#1S)}}\xspace}

\def\LorLbar     {\kern 0.18em\optbar{\kern -0.18em \PLambda}{}\xspace}

\newcommand{\decay}[2]{\mbox{\ensuremath{#1\!\to #2}}\xspace}         

\def\to                 {\ensuremath{\rightarrow}\xspace}

\def\CP                {{\ensuremath{C\!P}}\xspace}

\def\AT#1     {\ensuremath{A_{\mathrm{T}}^{#1}}\xspace}           

\def\C#1      {\ensuremath{\mathcal{C}_{#1}}\xspace}                       
\def\Cp#1     {\ensuremath{\mathcal{C}_{#1}^{'}}\xspace}                    
\def\Ceff#1   {\ensuremath{\mathcal{C}_{#1}^{\mathrm{(eff)}}}\xspace}        
\def\Cpeff#1  {\ensuremath{\mathcal{C}_{#1}^{'\mathrm{(eff)}}}\xspace}       
\def\Ope#1    {\ensuremath{\mathcal{O}_{#1}}\xspace}                       
\def\Opep#1   {\ensuremath{\mathcal{O}_{#1}^{'}}\xspace}                    

\newcommand{\tev}{\ifthenelse{\boolean{inbibliography}}{\ensuremath{~T\kern -0.05em eV}}{\ensuremath{\mathrm{\,Te\kern -0.1em V}}}\xspace}
\newcommand{\gev}{\ensuremath{\mathrm{\,Ge\kern -0.1em V}}\xspace}
\newcommand{\mev}{\ensuremath{\mathrm{\,Me\kern -0.1em V}}\xspace}
\newcommand{\kev}{\ensuremath{\mathrm{\,ke\kern -0.1em V}}\xspace}
\newcommand{\ev}{\ensuremath{\mathrm{\,e\kern -0.1em V}}\xspace}
\newcommand{\mevc}{\ensuremath{{\mathrm{\,Me\kern -0.1em V\!/}c}}\xspace}
\newcommand{\gevc}{\ensuremath{{\mathrm{\,Ge\kern -0.1em V\!/}c}}\xspace}
\newcommand{\mevcc}{\ensuremath{{\mathrm{\,Me\kern -0.1em V\!/}c^2}}\xspace}
\newcommand{\gevcc}{\ensuremath{{\mathrm{\,Ge\kern -0.1em V\!/}c^2}}\xspace}
\newcommand{\gevgevcc}{\ensuremath{{\mathrm{\,Ge\kern -0.1em V^2\!/}c^2}}\xspace} 
\newcommand{\gevgevcccc}{\ensuremath{{\mathrm{\,Ge\kern -0.1em V^2\!/}c^4}}\xspace} 

\def\gsim{{~\raise.15em\hbox{$>$}\kern-.85em
          \lower.35em\hbox{$\sim$}~}\xspace}
\def\lsim{{~\raise.15em\hbox{$<$}\kern-.85em
          \lower.35em\hbox{$\sim$}~}\xspace}

\def\tell1  {TELL1\xspace}
\def\ukl1   {UKL1\xspace}

\usepackage{cite} 
\usepackage{mciteplus}

\usepackage{bm}

\begin{document}

\renewcommand\Affilfont{\itshape\small}

\author[1]{Y.~Amhis}\affil[1]{LAL, Universit\'{e} Paris-Sud, CNRS/IN2P3, Orsay, France}
\author[2]{Sw.~Banerjee}\affil[2]{University of Louisville, Louisville, Kentucky, USA}
\author[3]{E.~Ben-Haim}\affil[3]{LPNHE, Sorbonne Universit\'e, Paris Diderot Sorbonne Paris Cit\'e, CNRS/IN2P3, Paris, France}
\author[4]{M.~Bona}\affil[4]{School of Physics and Astronomy, Queen Mary University of London, London, UK}
\author[5]{A.~Bozek}\affil[5]{H. Niewodniczanski Institute of Nuclear Physics, Krak\'{o}w, Poland}
\author[6]{C.~Bozzi}\affil[6]{INFN Sezione di Ferrara, Ferrara, Italy}
\author[5]{J.~Brodzicka}
\author[7]{M.~Chrzaszcz} \affil[7]{European Organization for Nuclear Research (CERN), Geneva, Switzerland}
\author[8]{J.~Dingfelder}\affil[8]{University of Bonn, Bonn, Germany}
\author[9]{U.~Egede}\affil[9]{Imperial College London, London, UK}
\author[10]{M.~Gersabeck}\affil[10]{School of Physics and Astronomy, University of Manchester, Manchester, UK}
\author[11]{T.~Gershon}\affil[11]{Department of Physics, University of Warwick, Coventry, UK}
\author[12]{P.~Goldenzweig}\affil[12]{Institut f\"ur Experimentelle Teilchenphysik, Karlsruher Institut f\"ur Technologie, Karlsruhe, Germany}
\author[13]{K.~Hayasaka}\affil[13]{Niigata University, Niigata, Japan}
\author[14]{H.~Hayashii}\affil[14]{Nara Women's University, Nara, Japan}
\author[7]{D.~Johnson}
\author[15]{M.~Kenzie}\affil[15]{Cavendish Laboratory, University of Cambridge, Cambridge, UK}
\author[16]{T.~Kuhr}\affil[16]{Ludwig-Maximilians-University, Munich, Germany}
\author[17]{O.~Leroy}\affil[17]{Aix Marseille Univ, CNRS/IN2P3, CPPM, Marseille, France}
\author[18,19]{H.~B.~Li}\affil[18]{Institute of High Energy Physics, Beijing 100049, People’s Republic of China}\affil[19]{University of Chinese Academy of Sciences, Beijing 100049, People’s Republic of China}
\author[20,21]{A.~Lusiani}\affil[20]{Scuola Normale Superiore, Pisa, Italy}\affil[21]{INFN Sezione di Pisa, Pisa, Italy}
\author[14]{K.~Miyabayashi}
\author[22]{P.~Naik}\affil[22]{H.H.~Wills Physics Laboratory, University of Bristol, Bristol, UK}
\author[23]{T.~Nanut}\affil[23]{Institute of Physics, Ecole Polytechnique F\'{e}d\'{e}rale de Lausanne (EPFL), Lausanne, Switzerland}
\author[9]{M.~Patel}
\author[24,25]{A.~Pompili}\affil[24]{Universita' di Bari Aldo Moro, Bari, Italy}\affil[25]{INFN Sezione di Bari, Bari, Italy}
\author[26]{M.~Rama}\affil[26]{INFN Sezione di Pisa, Pisa, Italy}
\author[27]{M.~Rotondo}\affil[27]{Laboratori Nazionali dell'INFN di Frascati, Frascati, Italy}
\author[23]{O.~Schneider}
\author[28]{C.~Schwanda}\affil[28]{Institute of High Energy Physics, Vienna, Austria}
\author[29]{A.~J.~Schwartz}\affil[29]{University of Cincinnati, Cincinnati, Ohio, USA}
\author[30,31]{B.~Shwartz}\affil[30]{Budker Institute of Nuclear Physics (SB RAS), Novosibirsk, Russia}\affil[31]{Novosibirsk State University, Novosibirsk, Russia}
\author[17]{J.~Serrano}
\author[32]{A.~Soffer}\affil[32]{Tel Aviv University, Tel Aviv, Israel}
\author[33]{D.~Tonelli}\affil[33]{INFN Sezione di Trieste, Trieste, Italy}
\author[34]{P.~Urquijo}\affil[34]{School of Physics, University of Melbourne, Melbourne, Victoria, Australia}
\author[35]{R.~Van Kooten}\affil[35]{Indiana University, Bloomington, Indiana, USA}
\author[36]{J.~Yelton}\affil[36]{University of Florida, Gainesville, Florida, USA}

\title{\papertitle \\ \large \paperauthors\\
\href{mailto:hflav-conveners@cern.ch}{hflav-conveners@cern.ch}}
\date{}
\maketitle

\begin{abstract}
  \noindent
    The Heavy Flavor Averaging Group provides with this document input to
    the European Strategy for Particle Physics. Research in heavy-flavor physics is an essential component of the particle-physics program, both within and beyond the Standard Model. To fully realize the potential of the field, we believe the strategy should include strong support for
    the ongoing experimental and theoretical heavy-flavor research, future upgrades of existing facilities, and significant heavy-flavor capabilities at future colliders, including dedicated experiments.
\end{abstract}

{\footnotesize 
\centerline{\copyright~\papercopyright. \href{\paperlicenceurl}{\paperlicence}.}}
\vspace*{2mm}

\renewcommand{\thefootnote}{\arabic{footnote}}
\setcounter{footnote}{0}


\parskip=2mm

\noindent 
The Heavy Flavor Averaging Group (HFLAV)~\cite{HFLAV-web} was formed in 2002 to 
continue the activities of the LEP Heavy Flavor Steering 
Group~\cite{Abbaneo:2001bv_mod_cont}. 
HFLAV is responsible for calculating world averages of measurements of 
beauty-hadron, charm-hadron and tau-lepton properties from current and 
past experiments, and provides a comprehensive resource for the field in terms of web pages 
and full documentation of results. The most recent compilation of our results appears in Ref.~\cite{HFLAV16}. Many of our world averages are used by the Particle Data Group. 
With this perspective, we take the opportunity to comment on the importance of future heavy-flavor physics research. \textbf{In this document, we make the argument for support of heavy-flavor physics in both the near term and the far future.} 

Flavor physics is among the central foundations of the Standard Model
(SM). Indeed, many advances in the construction of the SM originated
from research into flavor physics. This included the three-generation prediction of the
Kobayashi-Maskawa mechanism, the universality of the gauge interactions, the high
masses of the top quark and the weak gauge bosons, and the presence of
large charge-parity (\CP) violation in beauty hadrons. Similarly, heavy-flavor physics is
closely tied to questions that aim to unravel the physics beyond the SM
(BSM). Some of these questions, and the ways in which current research
in heavy-flavor physics studies them, are as follows:
\begin{itemize}
\item The baryon asymmetry of the universe necessitates \CP violation
  far beyond that provided by the SM. Precise measurements of \CP
  violation provide unique access to \CP violation in BSM
  physics. Processes in which the SM predicts zero or very small \CP
  violation can be particularly sensitive to BSM amplitudes. These
  include decays of the tau lepton and specific charm- and
  beauty-hadron decays.

\item The origins of the three generations of fermions and of the
  Yukawa couplings that distinguish between them lie in BSM
  physics. This is probed by precision studies of the
  Cabibbo-Kobayashi-Maskawa (CKM) matrix, which test the
  three-generation picture and the SM hypothesis that quark-flavor nonuniversality depends on only the four parameters of the CKM matrix.

\item The SM gauge interactions are independent of flavor. However, the
  flavor-nondiagonal Yukawa couplings motivate searching for
  flavor-nonuniversal BSM interactions, which may arise from new
  gauge bosons or scalars, and may also be related to the dark-matter puzzle.
  Thus, it is important to test universality in heavy-flavor decays. In particular, lepton-flavor nonuniversality, lepton-flavor violation and lepton-number violation are unambiguous and sensitive probes of BSM physics. 
  Similarly, flavor-changing neutral currents (FCNCs), which in the SM occur only
  at loop level, are very sensitive to the presence of heavy new states. 
\end{itemize}

The sensitivity of heavy-flavor measurements leads to tight constraints on
BSM physics, in many cases at energy scales that are far beyond
those accessible at energy-frontier facilities. Examples of recent heavy-flavor results that have received a great deal of attention in the phenomenology and model-building literature include measurements of lepton-flavor universality
and the observation of rare bottom-meson decays into $\mu^+\mu^-$. In turn, some of these measurements have motivated searches for new heavy mediators at the LHC~\cite{Sirunyan:2018kzh, *Sirunyan:2018jdk}. In at least some of the channels, studies show that only a 100~TeV collider will be able to cover most of the  model-parameter space allowed by the heavy-flavor constraints~\cite{Allanach:2017bta}.
Thus, \textbf{heavy-flavor
physics has far-reaching BSM sensitivity, as well as an important role 
in informing research at current and future energy-frontier facilities.} 

\textbf{Heavy-flavor physics provides a unique laboratory for studying the
strong interaction.} In particular, a number of hadrons
with nonstandard quantum numbers that contain charm or beauty quarks, such as the $X(3872)$ and the $Z_c^\pm(4430)$, 
have been discovered in the past 15 years. Revealing new ways in which QCD forms bound states, these discoveries have opened up a very active area of hadronic-physics research.

Heavy-flavor research is a major effort in Europe.
Across the continent, there are 61 groups on the \lhcb experiment, 36 on the
\belletwo experiment, and 17 on the \besiii experiment. In addition,
a significant number of groups work on heavy-flavor physics within  \atlas and
    \cms, and many European groups are leaders in the related theory.
As in other areas of particle physics, data analysis in heavy-flavor physics is 
often performed in small groups at universities and laboratories, involves detailed and specific collaboration with phenomenologists, and can 
take several years to complete due to its high complexity. For these reasons,
the long-term support of the experimental and theoretical groups that are involved in 
heavy-flavor research and train the students and postdocs is essential for the success of the field and for the efficient exploitation of the investment in the facilities. 
\textbf{A healthy research program requires support for CERN and 
other global laboratories, 
as well as the individual experimental and theoretical research
groups throughout European universities and institutes.}

In the past, successful heavy-flavor programs were carried out at the LEP experiments and SLD, the Tevatron experiments, and ARGUS and CLEO. BABAR and Belle dominated the field in the first decade of this century, and are still producing unique results. Currently, almost half the publications in  heavy-flavor physics originate from research at \lhcb, followed closely by \besiii and Belle, with fewer contributions from BABAR, \atlas and \cms. By far, the majority of beauty-hadron measurements in the next decade will come
from  \lhcb and \belletwo.
\lhcb benefits from large cross-sections for production of all types of beauty and charm hadrons and from precise timing that arises from the high boost of the produced particles. 
\belletwo will exploit production of $B$-meson pairs with well known kinematics in a clean environment, allowing better reconstruction of photons, particularly from $\pi^0$ and $\eta$ decays. To give an example of their complementarity in one case of interest, \lhcb will 
provide the most accurate measurements of angular distributions in exclusive electroweak-penguin decays with charged leptons and hadrons, while \belletwo will study
the corresponding inclusive decays and exclusive decays with soft photons, and will measure the branching fractions 
of the related decays $\decay{B}{K^{(\ast)}\nu\bar\nu}$. The combination of 
these results will constitute a stringent test of BSM physics involving 
new couplings to leptons and the $b$ quark. 
Similarly, in charm physics, \lhcb and \belletwo will produce the majority of 
high-statistics results, while BESIII, and possibly future charm-tau
factories that are being studied in Russia and China,
will exploit the $e^+e^-\to D^0 \bar 
D^0$ process to perform quantum-correlation measurements. By contrast, the 
physics of $\tau$ leptons will be dominated by \belletwo, and will offer unique probes of BSM physics. As an example, \belletwo will be sensitive to lepton-flavor violation in $\tau$ decays down to branching fractions of order $10^{-10}$.
\textbf{The dual approach of conducting heavy-flavor physics in both 
the controlled} $\bm{e^+ e^-}$ \textbf{environment and in the higher
statistics hadronic environment should be supported both in the short term 
and into the future.}

Each generation of flavor-physics experiments involves significant technological advances and large luminosity increases. BABAR, Belle, and \lhcb have pushed the boundaries of the heavy-flavor physics that could be
performed at $e^+ e^-$ colliders and in a hadron collider environment, respectively. Furthermore, \belletwo, which has just started, will collect 50 times more data than \belle by the middle of the next decade~\cite{Kou:2018nap}. On the same time scale, \lhcb will collect 5 times more data than its current sample. Beyond Run-4 of the LHC, the \lhcb Upgrade~II~\cite{LHCb-PII-Physics} is proposing to collect yet an order of magnitude more data,
increasing the explored BSM mass scale by close to a factor two relative to that of the Upgrade~I program. The
theoretical uncertainties related to the measurements are under control in most areas, while others require parallel developments in lattice QCD. The \belletwo experiment and SuperKEKB accelerator have also begun discussion of upgrade opportunities to extend the data sample beyond the currently planned 50~ab$^{-1}$, and possibly to introduce polarization on the electron beam.
\textbf{Support for the construction and full exploitation of \lhcb Upgrade~II and upgrades of \belletwo is important for probing the flavor structure of BSM physics at significantly higher mass scales. }

Heavy-flavor physics will continue to play this important role in the post-LHC era. 
Experiments at a future high-luminosity $e^+ e^-$ collider operating at the $Z$ resonance would perform unique heavy-flavor studies in specific channels. Maximising the detector's ability in this area would probably require charged-hadron identification capability. 
At a circular, high-energy $pp$ collider, heavy-flavor physics would be best studied with a dedicated experiment, along the lines of \lhcb, where the advantages of optimized detector and trigger have been demonstrated. \textbf{It is imperative that planning of new facilities include consideration of the ways in which heavy-flavor physics can be further explored.}

\textbf{Heavy-flavor physics helps drive the search for BSM physics and the understanding of the SM. This will continue to be the case in the foreseeable future. To fully realize the potential of heavy-flavor physics, we believe the European Strategy for Particle Physics should include strong support for the experimental and theoretical research in this area, as well as the development of future facilities.}

\setboolean{inbibliography}{true}
\bibliographystyle{LHCb}
\bibliography{standard}

\end{document}